\documentclass[10pt, conference, letterpaper]{IEEEtran}

\usepackage{cite}
\usepackage{amsmath,amssymb,amsfonts}
\usepackage{algorithmic}
\usepackage{graphicx}
\usepackage{textcomp}
\usepackage{xcolor}
\usepackage{xspace}
\usepackage{hyperref}
\usepackage{subcaption}
\usepackage{csquotes}
\usepackage{mathtools}
\usepackage{caption}

\newcommand{\Shortcut}{\textsc{Shortcut}\xspace}

\title{\Shortcut: In-Collective Topology Reconfiguration for Low-Latency AllReduce}

\begin{document}

\author{\IEEEauthorblockN{Sarah-Michelle Hammer}
\IEEEauthorblockA{\textit{Technische Universität Berlin} \\
Berlin, Germany}
\and
\IEEEauthorblockN{Anton Juerss}
\IEEEauthorblockA{\textit{Weizenbaum Institute} \\
\textit{\& Technische Universität Berlin}\\
Berlin, Germany}
\and
\IEEEauthorblockN{Stefan Schmid}
\IEEEauthorblockA{\textit{Technische Universität Berlin} \\
Berlin, Germany}
}

\maketitle

\begin{abstract}
Distributed ML training relies on efficient AllReduce communication to aggregate data across nodes. In this setting, reconfigurable optical interconnects offer high-bandwidth, energy-efficient direct links between accelerators but often produce ring-based topologies that remain static during a collective. 

The Ring AllReduce algorithm naturally matches these topologies but its cumulative latency grows linearly with node count. Low-latency algorithms such as Recursive Doubling (RD) instead achieve logarithmic cumulative per-step latency, but their long-distance exchanges incur dilation and congestion costs on a static ring. In-collective topology reconfiguration can eliminate these penalties but each topology change adds reconfiguration delay. The key question to improve AllReduce completion time is therefore not only how to reconfigure RD efficiently, but when selectively reconfigured RD becomes faster than the topology-matched Ring algorithm.

We present \Shortcut: an effective strategy for topology reconfiguration that enables RD to shortcut costly multi-hop communication only when it pays off~---~beyond the performance of the Ring algorithm. Across small and medium messages on 32 nodes, \Shortcut achieves $4.4\times$-$6.0\times$ speedups over Ring. At 128 nodes, it remains up to $7\times$ faster with a $10\,\mu s$ reconfiguration delay, showing that selective reconfiguration is especially effective in latency-sensitive, large-scale settings.
\end{abstract}

\section{Introduction}
\label{sec:introduction}

Collective communication operations such as AllReduce play a pivotal role in the performance of both distributed machine learning (ML) and high-performance computing applications~\cite{narayananEfficientLargescaleLanguage2021,zhangNetworkBottleneckDistributed2020,thakurOptimizationCollectiveCommunication2005}. The trend toward exponentially larger ML models and training datasets~---~requiring distribution across ever-larger accelerator counts~---~combined with faster individual accelerators, has made collective communication completion time a critical performance constraint~\cite{10.1145/3452296.3472900,klenk_InNetwork_2020,singhBigSendoffHigh2025}.

Optical circuit-switched (OCS) interconnects offer a promising path forward: by directly steering light paths they avoid the energy and scalability limitations of electrical packet processing~\cite{10.1145/3452296.3472900,ding_Photonic_2025,wu_actina_2025} and can reconfigure connectivity between accelerators to match the structured, repetitive communication patterns of ML training~\cite{wangTopoOptCooptimizingNetwork2023,10.1145/3452296.3472900,10664412}. Such circuit-switched topologies typically impose a low node degree, as each endpoint can typically terminate only a small number of direct connections and OCS switch radix is limited. The resulting ring-based topologies are reconfigured between jobs~\cite{wangTopoOptCooptimizingNetwork2023,jouppiTPUV4Optically2023,Lightwave_liu_2023} or communication phases~\cite{ding_Photonic_2025} but kept static during collective communication operations. 

On such static, direct-connect topologies, flows between non-adjacent nodes must traverse intermediate hops, incurring two key penalties: (i) the cumulative per-hop delay from propagation and processing, and (ii) congestion, as overlapping flows contend for shared links. 

Multi-hop penalties especially affect low-latency AllReduce algorithms that communicate between distant nodes, such as Recursive Doubling (RD). RD is generally considered latency-optimal because it completes AllReduce in $O(\log n)$ communication steps~\cite{sensi_swing_2024}, compared with the Ring Algorithm's $O(n)$ steps. Yet, we find that RD's multi-hop penalties can significantly outweigh its lower step count (§~\ref{sec:multihop}).

These limitations can be addressed by allowing the reconfigurable topology to evolve \emph{during} a collective. Establishing direct links between communicating nodes in each step can remove multi-hop delays and path-induced congestion, allowing RD to recover its lower-step-count advantage. Recent work has shown such on-demand links are feasible~\cite{10.1145/3696348.3696856,10.1145/3452296.3472900}, but a key challenge remains: reconfiguration of optical interconnects takes time, and repeatedly paying this delay can offset or outweigh the resulting latency benefit.

The opportunity therefore lies between two costly extremes: a static topology, where RD suffers high multi-hop costs, and reconfiguration at every step, where it incurs high reconfiguration delay. Not every RD step requires the same topology support, as some exchanges incur substantially larger multi-hop penalties than others. Selectively reconfiguring only these steps can allow more of RD's theoretical advantage to translate into collective completion time improvements.

Recent work on in-collective reconfiguration mainly asks how to optimize the reconfiguration schedule for a given collective communication algorithm~\cite{photonicscaleup,Harvest_rahman_2026,PCCL_kumar_2025}. Such general formulations require costly maximum concurrent-flow calculations or ILP optimizations and focusing on one algorithm alone leaves open the trade-off \emph{between algorithms}.

Selective reconfiguration can improve low-step-count algorithms like RD, but this does not necessarily make them faster than the Ring Algorithm. The Ring algorithm already fits the physical topology: it communicates only between direct neighbors and therefore pays neither multi-hop penalties nor reconfiguration delay. Reconfigured low-latency algorithms instead trade fewer communication steps against the cost of additional reconfiguration delay. The broader design space, and a clear understanding of the performance trade-offs between algorithms across reconfiguration delays, message sizes, and latencies, remains largely unexplored. This gap motivates our central question:

\smallskip
\emph{(When) can selective reconfiguration outperform an algorithm naturally matched to the underlying topology?}
\smallskip

We address this with \Shortcut: a threshold-based reconfiguration schedule for Recursive Doubling that carefully navigates the trade-off between reconfiguration benefit and cost and either chooses to reconfigure the topology for some steps of RD or falls back to a static ring topology and Ring algorithm~---~improving performance when possible, but never degrading it. This simplicity makes \Shortcut practical: the threshold calculation uses only static or locally available parameters known before the collective begins, and requires evaluating only $O(\log n)$ candidate thresholds, avoiding both explicit coordination and complex optimization to find the schedule.

We implement support for reconfigurable direct-connect topologies in the state-of-the-art ML training simulator Astra-Sim~\cite{rashidi2020astrasim} and its ns-3 network backend, and use it to compare \Shortcut directly against collective communication algorithms on the static topology. Across realistic optical reconfiguration delays, \Shortcut reduces collective completion time by up to $14.7\times$, with the largest gains in latency-dominated, large-scale settings. Our analysis and evaluation characterize the performance boundary between both approaches across message sizes and delays, showing when selective reconfiguration is beneficial and when a topology-matched Ring algorithm remains preferable. We will release our Astra-Sim/ns-3 extensions as open-source code.

\section{Background}\label{sec:background}
We first describe the reconfigurable optical interconnects considered in this paper and the topology constraints they impose on collective communication. We then introduce the cost model for AllReduce algorithms and its extension to capture these topology-dependent effects.

\subsection{Optical Reconfigurable Topologies for ML}

\begin{figure}
    \centering
    \includegraphics[width=\linewidth]{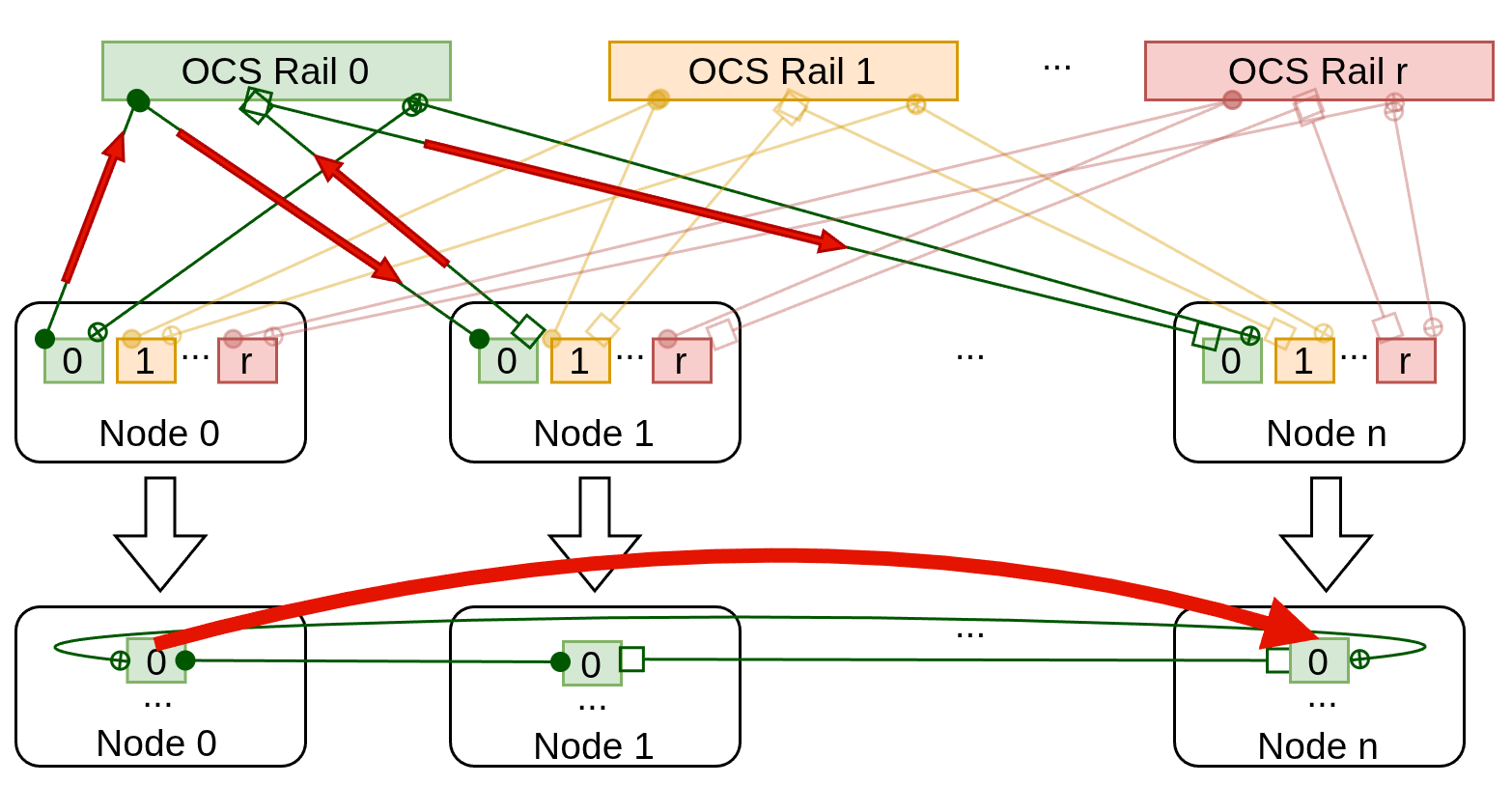}
    \caption{Photonic rails result in ring topologies between accelerators with the same index (rail). A flow between non-adjacent nodes in the ring (red arrow) has multiple hops across the photonic interconnect.}
    \label{fig:photonic_rail}
\end{figure}

Optical circuit-switching has been proposed to address the power and scalability limits of large packet-switched network architectures for ML~\cite{wangTopoOptCooptimizingNetwork2023,wu_actina_2025,ding_Photonic_2025,305352,10.1145/3452296.3472900,photonicscaleup, 10.1145/3696348.3696856}, and has already been deployed at scale in Google's TPUv4~\cite{jouppiTPUV4Optically2023}. We model such an optical circuit-switched interconnect as a reconfigurable direct-connect network: at any point in time, the circuit switch establishes a set of point-to-point links between nodes by steering optical paths, rather than electronic processing and forwarding at intermediate switches. In this way, the OCS configuration defines the edge set between nodes.

\begin{table}
    \centering
    \caption{Reconfiguration time and port count trade-off demonstrated
    by optical circuit-switching prototypes.}
    \label{tab:ocs_prototypes}
    \setlength{\tabcolsep}{3.5pt}
    \renewcommand{\arraystretch}{1.08}
    \begin{tabular}{lcc}
        \hline
        \textbf{OCS Technology}
        & \textbf{Reconfiguration Time}
        & \textbf{Port Count} \\
        \hline
        Wavelength-routed~\cite{ballani_Sirius_2020}
        & 100~ns
        & 8 \\
        Silicon Photonics, in~\cite{ding_Photonic_2025}
        & $7 \mu s$ & 32 \\
        Rotor switch~\cite{Realizing_mellette_2024}
        & 7~$\mu$s & 128 \\
        3D MEMS, in~\cite{ding_Photonic_2025} & 15~ms  & 320\\  
        \hline
    \end{tabular}
\end{table}

Existing architecture proposals and deployments typically reconfigure the optical topology at a coarser granularity: before a job begins~\cite{wangTopoOptCooptimizingNetwork2023,jouppiTPUV4Optically2023,
10.1145/3452296.3472900}, or between communication phases ~\cite{ding_Photonic_2025,wu_actina_2025}.  Within an individual collective,
however, the resulting direct-connect topology generally remains static.

Photonic rails illustrate this setting. Accelerators belonging to the same rail (same server-local index) are connected in a direct-connect ring through links established by the OCS, as illustrated in Fig.~\ref{fig:photonic_rail}. Adjacent accelerators communicate directly, whereas communication between non-adjacent nodes must traverse multiple hops through the interconnect and intermediate endpoints.

In the remainder of this paper, we use \emph{static ring} to refer to such a topology held fixed for the duration of a collective, and \emph{reconfiguration} to refer to changing the circuit-switched matching between collective steps.

\subsection{Collective Communication Cost Model}
\label{sec:classic_cost_model}

\textit{Collective communication operations}, such as AllReduce, specify the
transformation from input to output buffers across all participating nodes but do not prescribe how data is exchanged. A \textit{collective communication
algorithm} specifies the concrete point-to-point flows and their
dependencies~\cite{thakurOptimizationCollectiveCommunication2005}. These dependencies organize flows into sequential steps: flows in a step can execute in parallel, but each step must complete before the next begins. 

The performance of collective algorithms is commonly analyzed using the
$\alpha$--$\beta$ cost model~\cite{thakurOptimizationCollectiveCommunication2005}, which abstracts the physical topology into a fixed per-step startup latency $\alpha_s$ and a per-byte transmission time $\beta$. The  Ring and Recursive Doubling (RD) algorithms both implement AllReduce as a concatenation of ReduceScatter followed by AllGather. For an AllReduce message of size $m$ across $n$ nodes their costs are\cite{thakurOptimizationCollectiveCommunication2005}
\begin{align}
    t_{\alpha-\beta}(\mathrm{Ring}) &= 2\bigl((n-1)\alpha_s + \frac{n-1}{n}m\beta\bigr) \nonumber \\ t_{\alpha-\beta}(\mathrm{RD}) &= 2\bigl((\log_2 n)\alpha_s + \frac{n-1}{n}m\beta\bigr) \nonumber.
\end{align}
Since the $\beta$-terms of both algorithms are identical, RD's advantage lies in its $O(\log n)$ startup latency versus Ring's $O(n)$. Algorithms such as RD are thus regarded as \emph{low-latency} collective algorithms~\cite{singhBigSendoffHigh2025,thakurOptimizationCollectiveCommunication2005}. Similarly, communication is \emph{latency-bound} when the $\alpha_s$ term dominates the total cost, which commonly occurs for small message sizes or large node counts. The bandwidth term $\frac{n-1}{n} \; m \beta \approx m \beta$ is nearly independent of $n$, so as the node count grows, latency increasingly dominates total cost. 

In practice, large-scale ML training deployments face increased latency pressure. They experience significantly higher latencies due to multi-hop paths and long cable runs (up to $30\times$ higher than intra-rack latencies~\cite{Collective_si_2025}). Larger node counts, e.g. 128 nodes for the data-parallel dimension in Llama model training~\cite{Llama_grattafiori_2024}, make the linear startup latency of Ring particularly costly. This has driven adoption of the Recursive Doubling algorithm~---~long established in the HPC community~\cite{thakurOptimizationCollectiveCommunication2005}~---~in large-scale ML deployments, to mitigate latency pressure at high node counts~\cite{Collective_si_2025}.

\subsection{Topology-aware Cost Model}
\label{sec:topology-aware cost model}
\begin{figure}[t]
    \centering
    \includegraphics[width=0.9\linewidth]{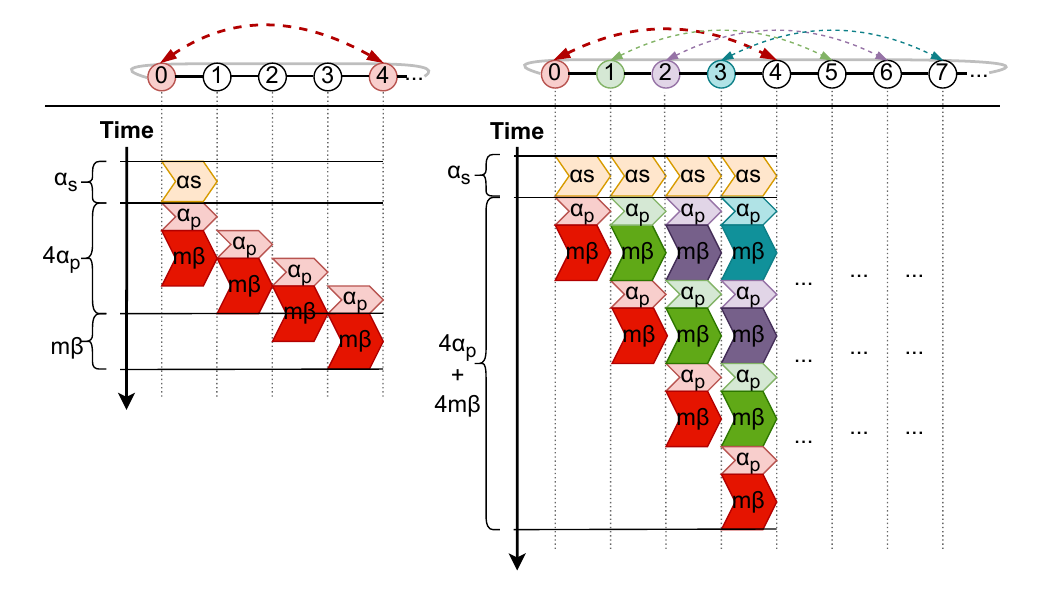}
        \caption{Per-hop delay ($\alpha_p$) and congestion (multiple $m\beta$) penalties of multi-hop flows on ring topologies. 
        Flow order does not affect the maximum flow completion time and hence the collective completion time.}
    \label{fig:multi-hop-flow-penalities}
 \end{figure}
 
The conventional $\alpha$--$\beta$ model does not capture how a communication
pattern maps onto a reconfigurable direct-connect topology. We therefore use a more general per-step model that makes path length, congestion, and topology reconfiguration explicit, based on~\cite{sensi_swing_2024,juerss2026,photonicscaleup} and illustrated in Figure~\ref{fig:multi-hop-flow-penalities}. For the cost of communication step $i$, we write
\begin{equation}
\label{eq:general-step-cost}
t_c^i = \alpha_s \,+\, h_i \, \alpha_p\,+\, c_i \, m_i\beta\,+\,r_i \, \delta. 
\end{equation}
where $\alpha_p$ is the per-hop delay, scaled by the maximum hop count $h_i$ in step $i$, $m_i$ is the maximum flow size in this step and $c_i$ models the bandwidth reduction caused by congestion. This captures reduced bottleneck bandwidth under contention of multiple flows, akin to congestion deficiency in~\cite{sensi_swing_2024}. For example, two flows sharing a link are modeled as $2\beta m$ because it takes twice as long to transmit the message on the bottleneck link. Lastly, $r_i \in \{0,1\}$ indicates whether the OCS is reconfigured before this step, adding a reconfiguration delay $\delta$. This is summed across all steps of a collective communication algorithm to calculate its total cost.

\section{The Case for In-Collective Reconfiguration}\label{sec:motivation}
OCS-based direct-connect topologies favor collective algorithms that match their limited connectivity, but can suppress the latency advantage of algorithms with more distant communication patterns. We next show why ring-based designs are a practical operating point, why Ring AllReduce is difficult to outperform on them, and where in-collective reconfiguration creates an opportunity.

\subsection{Rings as an Emerging Reconfigurable Topology}
With one port per node, an OCS with $(p = n)$ endpoint-facing ports can realize only a matching, so the direct-connect topology is disconnected for $n>2$ and must be reconfigured between communication steps to allow communication between different node pairs. A direct-connect topology that can remain connected during all steps of a collective, while still allowing links to be reconfigured between steps, therefore requires at least two ports per node, or $(p=2n)$ OCS ports in total. The corresponding minimum, uniform-degree connected topology is a ring which arises frequently in cost-effective HPC and distributed ML system designs~\cite{sensi_swing_2024, AmazonEC2Trn2} as well as in optical circuit-switched interconnect designs for ML~\cite{10.1145/3452296.3472900, wangTopoOptCooptimizingNetwork2023,jouppiTPUV4Optically2023,ding_Photonic_2025,infiniteHBD}. 

Higher endpoint degree can increase aggregate bandwidth and reduce topology diameter, but it also requires more OCS ports to realize these edges. Given the OCS radix-delay trade-off summarized in Table~\ref{tab:ocs_prototypes}, the corresponding longer reconfiguration delay increases total completion time and constrains feasible reconfiguration schedules. Therefore, we first study the common $(p=2n)$ configuration, which maximizes the number of nodes accommodated by a single OCS and extend this design to $D$-dimensional tori with $p=2nD$ ports. 

\subsection{Ring AllReduce Is Difficult to Beat}
\label{sub:ringRD}
Under the conventional $\alpha$--$\beta$ model (§\ref{sec:classic_cost_model}), Ring and RD should have the same bandwidth term and should therefore have comparable completion times when startup latency is negligible. As captured by the topology-aware model, however, this comparison changes once their communication patterns are mapped onto the physical topology. 

\begin{figure}[t]
\centering
\includegraphics[width=0.75\linewidth]{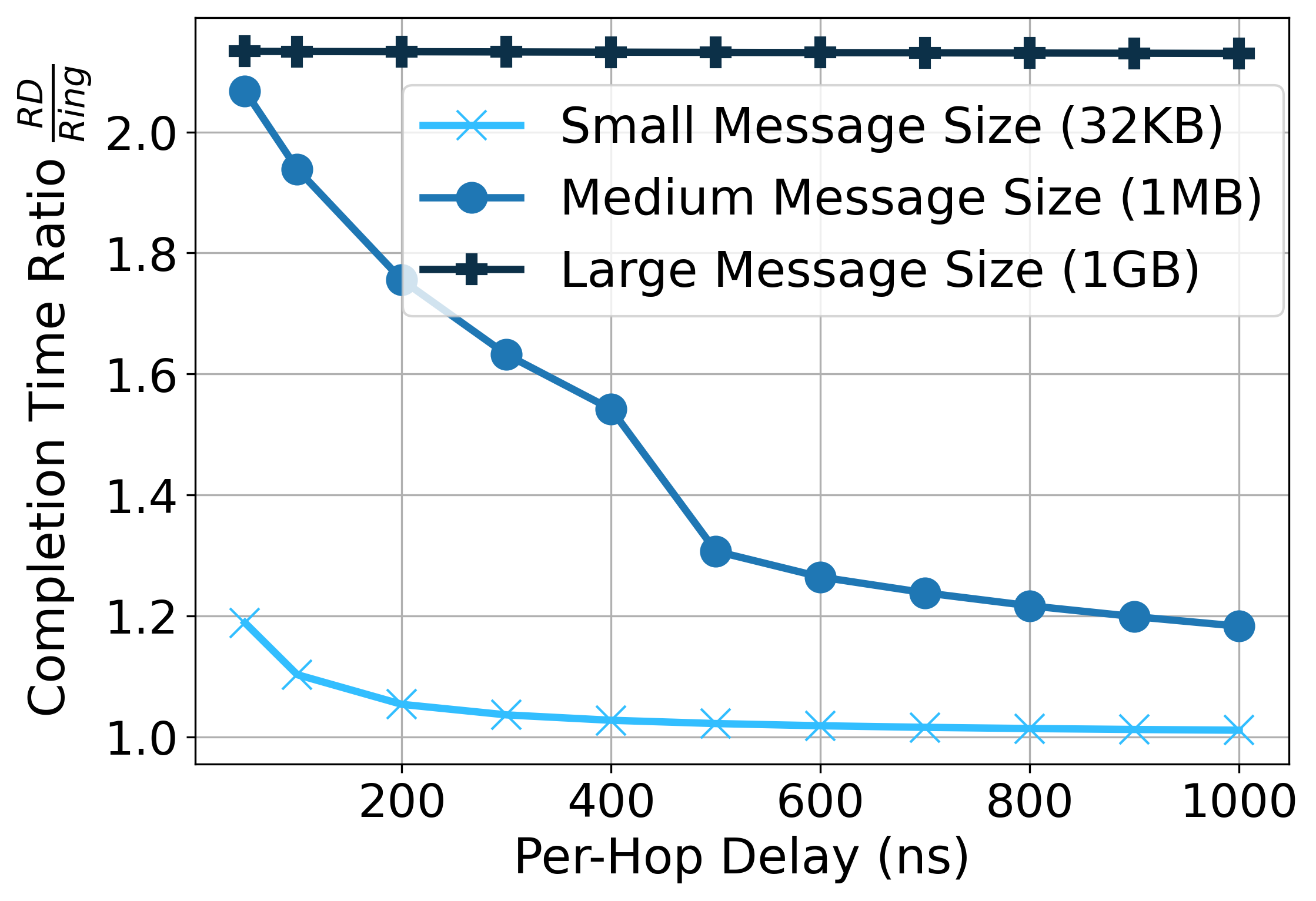}
\caption{Ring AllReduce outperforms Recursive Doubling (Ratio $>1$) even for small message sizes.}
\label{fig:ring-recursive-doubling}
\end{figure}

We illustrate this effect by comparing Ring and RD AllReduce on a static ring topology using Astra-Sim~\cite{rashidi2020astrasim} with ns-3. We model 16 GPUs connected by 800~Gbps links, vary per-hop delay and message size, and assume negligible startup latency $\alpha_s = 0$.

Contrary to the conventional model, Fig.~\ref{fig:ring-recursive-doubling} shows that Ring consistently outperforms RD across all evaluated message sizes from 1~KB to 1~GB. Even for 32~KB messages, RD is at least $1.2\times$ slower, and the gap widens as per-hop delay increases. This performance gap results from RD's increasingly long and overlapping multi-hop paths on the static ring, which accumulate propagation and processing delay and create congestion on shared links, substantially changing the relative performance of both algorithms.

\subsection{The Opportunity: Latency Optimization}
Ring performs well because its communication pattern naturally matches the topology. Every flow uses a direct neighboring link, so the algorithm avoids multi-hop dilation and congestion. Ring is also bandwidth-optimal for AllReduce in the sense that it transfers the minimal amount of data required by the operation~\cite{sensi_swing_2024}. Consequently, there is little opportunity to improve AllReduce completion time by reducing data movement or avoiding congestion. The main remaining opportunity is to reduce Ring's cumulative latency, which grows linearly due to its $2(n-1)$ communication steps.

This opportunity becomes increasingly important at scale. As discussed in Section~\ref{sec:background}, larger node numbers make Ring's cumulative startup latency more costly, while long physical paths and per-hop processing further increase latency pressure. A low-step-count algorithm such as RD offers an appealing alternative because it can reduce these latency overheads.

\subsection{Static Topologies Miss Low-Latency Advantages}\label{sec:multihop}
On direct-connect ring topologies, the limited degree of connectivity imposes a fundamental constraint on low-latency algorithms: the distance that can be covered by direct, single-hop exchanges in $O(\log n)$ steps is smaller than the diameter of the ring network $\lfloor n/2 \rfloor \in O(n)$ as n grows. To distribute data among all $n$ participants in $O(\log n)$ steps, at least some communication must occur between non-adjacent nodes~--~regardless of the specific communication pattern of the algorithm. Hence, while refined communication patterns such as Swing~\cite{sensi_swing_2024} can lessen the impact of multi-hop penalties, they cannot fully eliminate the underlying constraints imposed by the static topology.

This opens a complementary opportunity: rather than adapting only the collective algorithm to the topology, adapt the topology to the algorithm. Establishing direct links for RD's most expensive multi-hop steps can recover part of its low-step-count advantage, but each topology change incurs reconfiguration delay. The key design question in order to efficiently reconfigure RD is therefore which steps are costly enough that their avoided multi-hop penalties outweigh the reconfiguration cost.

\section{Navigating Reconfiguration Delay}\label{sec:main}
We consider a reconfigurable architecture with an end-to-end reconfiguration delay $\delta$ connecting $n$ accelerator endpoints, each equipped with two communication ports. On this topology \Shortcut operates on a spectrum between two costly extremes for Recursive Doubling: keeping the ring topology static avoids reconfiguration but suffers severe multi-hop penalties, while reconfiguring between every step of RD removes these penalties at the cost of repeated reconfiguration delay. We first model the cost of both extremes and derive a threshold schedule that reconfigures only the most costly steps. The resulting schedule is compared with topology-matched Ring costs to identify the best strategy to improve AllReduce completion time.

\subsection{Static and Fully Reconfigured RD}\label{sec:modeling_multi-hop}
We explicitly model the multi-hop penalties of RD on ring topologies. We first instantiate Eq.~\ref{eq:general-step-cost} for RD ReduceScatter because RD AllGather follows the same pairwise communication steps in reverse, and RD AllReduce concatenates both phases.

The Recursive Doubling ReduceScatter algorithm progresses in $\log_2 n$. In each step~$i$, the algorithm halves the chunk size and doubles the communication distance between communicating pairs. Concretely, in a ring network topology, each communication step~$i$ (0-indexed) incurs a fixed startup latency $\alpha_s$ and a cumulative per-hop delay of $\alpha_p \cdot 2^{i}$, since the distance doubles with each step. 

In addition, each step~$i$ incurs a transmission delay of $\beta \cdot \frac{m}{2^{i+1}} \cdot 2^{i}$. The term $\frac{m}{2^{i+1}}$ is the chunk size sent in step~$i$, while the factor $2^{i}$ reflects the extra congestion: multiple overlapping chunks share the same link in that step. Essentially, the available bandwidth per chunk shrinks by a factor of $2^{i}$.

Putting this together, the completion time for each step~$i$ is:
\begin{equation}\label{eq:rd-step-cost}
t_{\text{static}}^i(\text{RD}) = \alpha_p \cdot \overbrace{2^i}^{\text{path length}} +\; \alpha_s +  \beta \cdot \frac{m}{2^{i+1}} \cdot \overbrace{2^{i}}^{\text{congestion}}
\end{equation}

Summing across all $\log_2 n$ steps of ReduceScatter gives the total completion time for Recursive Doubling:
\begin{equation}\label{eq:rd-total-cost}
t_{\text{static}}(\text{RD}) = \alpha_p \cdot (n - 1) + \alpha_s \cdot \log_2 n + \frac{\log_2 n}{2} m \beta.
\end{equation}

At the opposite extreme, reconfiguring the topology before every step directly connects each communicating pair. Thus, $h_i=c_i=1$ and $r_i=1$, producing
\begin{equation}
\label{eq:rd-fully-reconfigured-cost}
t_{\text{reconf.}}(\mathrm{RD}) = (\alpha_s+\alpha_p+\delta)\log_2 n + \frac{n-1}{n}m\beta .
\end{equation}

For comparison, Ring ReduceScatter performs $n-1$ direct and congestion-free steps, each carrying $m/n$ bytes:
\begin{equation}\label{eq:ring-total-cost}
t_{\text{static}}(\text{Ring}) =  \alpha_p \cdot (n - 1) +  \alpha_s \cdot (n - 1)  + \frac{n - 1}{n} m \beta.
\end{equation}

Comparing Eq.~\ref{eq:rd-total-cost} and~\ref{eq:ring-total-cost} reveals an important insight for static topologies: the per-hop delay term for both algorithms adds up to the same $\alpha_p \cdot (n - 1)$. However, the transmission delay term grows with a factor of $\frac{\log_2 n}{2}$ for Recursive Doubling due to congestion, while it stays close to $1$ for Ring (since $\frac{n - 1}{n} \approx 1$). When $\alpha_s$ is small, this simple yet crucial difference explains why the Ring algorithm remains superior even for small message sizes on ring topologies (as in §\ref{sub:ringRD}): it avoids long paths and the resulting congestion that make Recursive Doubling surprisingly expensive under realistic physical constraints.

\subsection{Threshold-Based Reconfiguration}
Each step in Recursive Doubling induces a perfect matching and hence maps naturally to an OCS matching, enabling direct links to be established on the fly: short-cutting the ring via reconfiguration. By balancing reconfiguration delays against reduced propagation and congestion costs, our approach dynamically decides when reconfiguration is worthwhile, achieving completion times that outperform both static RD and Ring.

Instead of reconfiguring between every step, \Shortcut uses the static ring up to a threshold step $T$. For all steps $i \geq T$, we reconfigure the topology to match the communication pattern of that step exactly. Since path length and congestion costs increase with step index (both $2^i$), once reconfiguration becomes beneficial, it remains so for subsequent steps, allowing a single threshold to govern the schedule.

We then compare the total completion time of Recursive Doubling with this circuit-switching schedule using $T$ against the Ring algorithm on a static topology. Our goal is to ensure that circuit-switching never hurts performance compared to the fallback. As such, we select the smallest $T$ satisfying:
\begin{align}
    & \underbrace{\sum_{i=0}^{T-1} \Big( \alpha_p \cdot \overbrace{2^i}^{\substack{\text{path}\\\text{length}}} + \alpha_s + \beta \frac{m}{2^{i+1}}\cdot \overbrace{2^i}^{\text{congestion}} \Big)}_{\text{Static ring until step } T} \label{eq:reconfiguration_vs_ringreducescatter} \nonumber \\
    &+\; \underbrace{\sum_{i=T}^{\log_2 n - 1} \Big( \alpha_p + \alpha_s + \delta + \beta \frac{m}{2^{i+1}} \Big)}_{\text{Per-step circuit-switching after step } T}  \nonumber \\
    &\le \underbrace{(\alpha_p + \alpha_s) \cdot (n - 1)+ \beta \cdot m \cdot \frac{n - 1}{n}}_{\text{Ring ReduceScatter on static ring}}
\end{align}

The AllGather phase applies the same technique in reverse, with threshold
$T'$ denoting the step after which we revert to the static ring:
\begin{align}
    &\underbrace{
    \sum_{i=0}^{T'-1}
    \Big(
      \alpha_p + \alpha_s + \delta + \beta \frac{m\cdot 2^i}{n}
    \Big)
  }_{\text{Per-step circuit-switching}} \nonumber \\
  &+\;
  \underbrace{\sum_{i=T'}^{\log_2 n - 1}
      \Big(
        \alpha_p \cdot \overbrace{2^{\log_2 n - i}}^{\substack{\text{path}\\\text{length}}}
        + \alpha_s + \beta \frac{m\cdot 2^i}{n} \cdot \overbrace{2^{\log_2 n - i}}^{\text{congestion}}
      \Big)
    }_{\text{Static ring after step } T'}
  \nonumber \\
  &\le
  \underbrace{
    (\alpha_p + \alpha_s)\cdot (n - 1) + \beta \cdot m \cdot \frac{n - 1}{n}
  }_{\text{Ring AllGather on static ring}}
    \label{eq:reconfiguration_vs_ring}
\end{align}

Finally, we simply iterate over all possible $T$ and $T^\prime$ (from $0$ to $\log_2 n - 1$) before execution and select the smallest $T$ that satisfies the inequality. If no $T$ exists that satisfies the inequality, we safely fall back to the Ring algorithm on the static topology. Given that the number of steps in Recursive Doubling is limited to $2\log_2 n$, our search space for $T$ is also limited to $O(\log n)$ values, making it efficient to compute. All required parameters ($\alpha_s$, $\alpha_p$, $\delta$, $m$, $n$) are locally available before execution, requiring no centralized coordination to calculate the schedule.

\section{Practical Realization of \Shortcut}
\label{sec:realization}

\Shortcut requires one fixed and one reconfigurable link per endpoint. For a power of two network of size $n$, the fixed links realize the matching
\begin{equation}
    M_{\mathrm{fix}}
    =
    \{\{j,j+1\}\mid j \bmod 2 = 0\}.
\end{equation}
The reconfigurable ports connected to the dynamic OCS with $p=n$ endpoint-facing ports have an initial configuration of
\begin{equation}
    M_{\mathrm{ocs}}
    =
    \{\{j,j+1\}\mid j \bmod 2 = 1\}.
\end{equation}
The union of both matchings $M_{\mathrm{fix}}\cup M_{\mathrm{ocs}}$ forms the ring required to run Recursive Doubling until threshold $T$. 

Following related work on hybrid optical topologies~\cite{griner2021cerberus, Farrington2010Helios, jouppiTPUV4Optically2023}, the fixed matching may be realized using dedicated electrical links, constructing a hybrid architecture with one electrical and one reconfigurable port per endpoint, reducing the required OCS radix for in-collective reconfigurations from $p=2n$ to $p=n$.
These links should provide a similar bandwidth to the optical connections. Alternatively, the fixed matching may be implemented by a separate, potentially larger OCS with higher reconfiguration delay that is reconfigured only between workloads and remains static during each collective. Its reconfiguration delay therefore does not affect the in-collective schedule, and its ports do not contribute to the radix requirement of the latency-critical dynamic OCS. Otherwise, both endpoint ports connect to a single OCS with $p=2n$, although the larger switch radix may increase reconfiguration delay. In all cases, the dynamic OCS with $p=n$ ports performs the in-collective reconfigurations required by \Shortcut.

During step $i$, the dynamic OCS realizes the communication matching of the pairwise collective algorithm between a node $u$ and its peer. For Recursive Doubling, this translates to
\begin{equation}
\label{math_RD}
    M_i
    =
    \{\{u,u\mathbin{\mathrm{xor}}2^i\}
      \mid 0 \leq u < n\}.
\end{equation}
Each Recursive Doubling step induces a perfect matching. For $i=0$, this matching coincides with $M_{\mathrm{fix}}$, so all communicating pairs are connected directly through the fixed fabric. For steps $1\leq i<T$, multi-hop transmissions use the static topology formed jointly by the fixed fabric and the dynamic OCS configuration.

An OCS must connect every endpoint within the same transitive communication group, i.e., all endpoints that may communicate either directly or through a sequence of exchanges. Step $i=0$ is realized entirely by the fixed matching $M_{\mathrm{fix}}$, therefore the two endpoints do not need to be attached to the same dynamic OCS. For every subsequent step $i\geq 1$, endpoint $u$ communicates with $u\mathbin{\mathrm{xor}}2^i$. Since $2^i$ does not modify the least-significant bit for $i\geq 1$, parity is preserved:
\begin{equation}
    (u\mathbin{\mathrm{xor}}2^i)\bmod 2 = u\bmod 2.
\end{equation}
Thus, after the first step, even-indexed endpoints communicate exclusively with even-indexed endpoints, and vice-versa for odd-indexed endpoints. The dynamic fabric can therefore be partitioned into two independent OCSes, one serving each parity class, with $p=\frac{n}{2}$ endpoint-facing ports per OCS. This preserves all dynamic Recursive Doubling matchings while reducing the required radix of each latency-critical OCS from $p=n$ to $p=\frac{n}{2}$.

\subsection{Extension to Multi-Dimensional Tori}

We extend the architecture from a single ring to a $D$-dimensional torus with dimension sizes $d_1,\ldots,d_D$, where each $d_k$ is a power of two and $n=\prod_{k=1}^{D}d_k$. Each endpoint is identified by a coordinate vector $x=(x_1,\ldots,x_D)$, with $0\leq x_k<d_k$. For the global identifier of $x$, $g_k$ denotes the index multiplier associated with dimension $k$. In particular, $g_1=1$, while each subsequent multiplier equals the product of the sizes of all preceding dimensions
\begin{equation}
\mathrm{id}(x)
=
\sum_{k=1}^{D} g_k x_k,
\qquad
g_1=1,
\qquad
g_k=\prod_{r=1}^{k-1} d_r
\quad \text{for } k\geq 2.
\end{equation}
For a single pairwise collective in a torus topology, each communication step forms a matching and can therefore be realized using one reconfigurable optical port per endpoint. However, \Shortcut executes the initial steps $i<T$ over the static torus, which requires each endpoint to maintain two incident links per dimension and thus a total degree of $2D$. We therefore provision one fixed and one reconfigurable port per endpoint and dimension. Their union forms the initial torus, while the reconfigurable ports subsequently establish direct links between communicating peers. The dynamic links by $M_{\mathrm{ocs}}$ in each dimension can additionally support mirrored collective schedules that independently communication along each of the $D$ dimensions. 

For each dimension $k$, the fixed fabric pairs consecutive endpoints along every one-dimensional ring parallel to that dimension. Let $\boldsymbol{e}_k$ denote the unit vector whose $k$'th entry is $1$ and remaining $0$. The fixed matching in dimension $k$ is
\begin{equation}
    M_{\mathrm{fix}}^{k}
    =
    \left\{
        \left\{
            \mathrm{id}(x),
            \mathrm{id}(x+e_k)
        \right\}
        \,\middle|\,
        x_k \bmod 2 = 0
    \right\}.
\end{equation}
By expressing the matching $M_{\mathrm{fix}}^{k}$ with global node indices, increasing the $k$th coordinate by one increases the global node identifier by the index multiplier $g_k$, i.e., $    \mathrm{id}(x+e_k)
    =
    \mathrm{id}(x)+g_k,
$. The fixed matching can therefore be written as

\begin{equation}
\begin{split}
    M_{\mathrm{fix}}^{k}
    &=
    \Bigl\{
        \bigl\{
\mathrm{id}(x),\mathrm{id}(x)+g_k
        \bigr\}
        \ \Bigm| \\
        &0\leq \mathrm{id}(x)<n,\;
\mathrm{id}(x)\bmod(2g_k)<g_k
    \Bigr\}.
\end{split}
\end{equation}
This condition selects the first block of size $g_k$ within every pair of consecutive blocks in dimension $k$. Adding $g_k$ then connects each selected endpoint to the endpoint at the subsequent coordinate in dimension $k$. Together, the fixed matchings form disjoint $D$-dimensional hypercubes, each containing $2^D$ endpoints. The dynamic OCS initially connects these hypercubes into a $D$-dimensional torus by establishing the remaining consecutive endpoint pairs, including the wrap-around edge of each ring. The initial OCS matching in dimension $k$ is
\begin{equation}
M_{\mathrm{ocs}}^{k}
=
\{
\{
\mathrm{id}(x),
\mathrm{id}(x+e_k)
\}
\mid
x_k \bmod 2 = 1
\}.
\end{equation}
For each communication step $i \geq T$, the matching $M_{\mathrm{ocs}}^{k}$ is replaced by the matching induced by the collective algorithm. For Recursive Doubling, this matching is given in Eq.~\ref{math_RD}.

This architecture closely resembles the block-based organization of Google’s TPUv4 system, in which $4\times4\times4$ groups of chips are connected internally through fixed electrical links and interconnected through optical circuits~\cite{jouppiTPUV4Optically2023}. The main distinction is that TPUv4 confines optical reconfiguration to connections between blocks, whereas our design exposes one fixed and one reconfigurable link per endpoint and torus dimension. The performance of \Shortcut in a torus is governed primarily by the size of the dimension along which the collective is executed. A collective along dimension $k$ operates over groups of $d_k$ endpoints and induces the same communication pattern, path lengths, and reconfiguration schedule as a ring collective with $d_k$ endpoints. Consequently, the relative speedup of \Shortcut along dimension $k$ matches that of a ring with $d_k$ endpoints. For example, peers communicate in a $16\times16$ torus only along a single dimension and therefore exhibit the same relative performance as \Shortcut on a ring with $n=16$.

\subsection{OCS Partitioning and Radix Trade-offs}

Extending the fixed--dynamic matching architecture from a ring to a $D$-dimensional torus substantially reduces the connectivity required from each dynamic OCS. As in the one-dimensional case, all Recursive Doubling steps $i\geq 1$ preserve the parity of $x_k$. Each dimension-$k$ ring can therefore be served by two independent dynamic OCSes, one for the even and one for the odd coordinate class, with $\frac{d_k}{2}$ endpoint-facing ports each. The minimum-radix assignment for dimension $k$ consequently requires 
$
2\prod_{r\neq k}d_r
=
\frac{2n}{d_k}
$ 
dynamic OCSes, each with $\frac{d_k}{2}$ ports. For an equal-sized torus with $d_k=d=n^{1/D}$ for every dimension, this yields OCSes with 
$
\frac{d}{2}
=
\frac{n^{1/D}}{2}
$ 
ports and a total of 
$
\sum_{k=1}^{D}\frac{2n}{d_k}
=
2D n^{\frac{D-1}{D}}
$ 
dynamic OCSes. This assignment minimizes the radix of each latency-critical OCS, which enables selecting ones with low reconfiguration delay. However, it is possible that multiple independent rings may instead share a larger OCS. If $b_k$ rings in dimension $k$ are grouped per parity class and connected through the same OCS, the architecture requires
$
\frac{2n}{b_k d_k}
$
OCSes with $\frac{b_kd_k}{2}$ ports each, assuming that $b_k$ divides $\frac{n}{d_k}$. Hence, the architecture exposes a direct trade-off between switch count and switch radix: assigning one ring per OCS minimizes radix, whereas consolidating rings reduces the number of OCSes at the cost of larger, potentially slower-switching devices.

\section{Evaluation}\label{sec:evaluation}
We structure our evaluation in three stages. First, we isolate the effect of reconfiguration on Recursive Doubling and ask: \emph{when and for which steps does reconfiguration improve RD?} This reveals how \Shortcut's threshold adapts to the competing costs of multi-hop communication and topology reconfiguration. Second, we ask whether the resulting selectively reconfigured \Shortcut algorithm correctly improves upon strong alternatives that retain a static topology: the topology-matched Ring algorithm and Swing, a low-latency AllReduce algorithm which adapts the communication pattern to reduce topology-induced penalties. Finally, we characterize the system configurations in which \Shortcut is most beneficial, considering startup latency, node number and the reconfiguration delays demonstrated by existing OCS prototypes.

\begin{figure}
\centering
\includegraphics[width=0.7\linewidth]{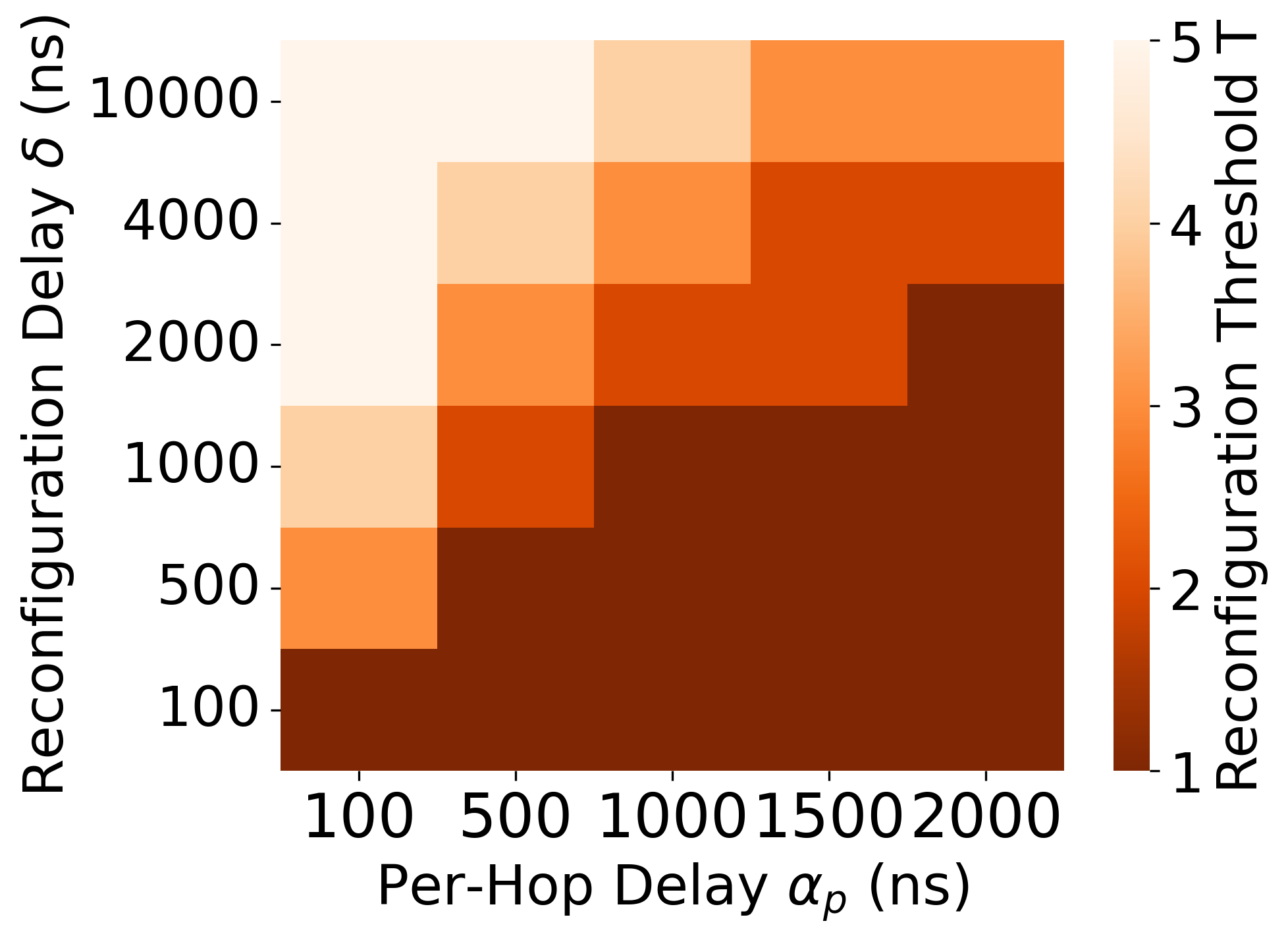}
\caption{Best reconfiguration threshold $T$ for 32~KB ReduceScatter at
         32~nodes. Lower, i.e. earlier, threshold $T$ when reconfiguration is fast or per-hop delay is high.}
\label{fig:heatmap_best_T}
\end{figure}

\begin{figure*}[th]
\centering
\includegraphics[width=\linewidth]{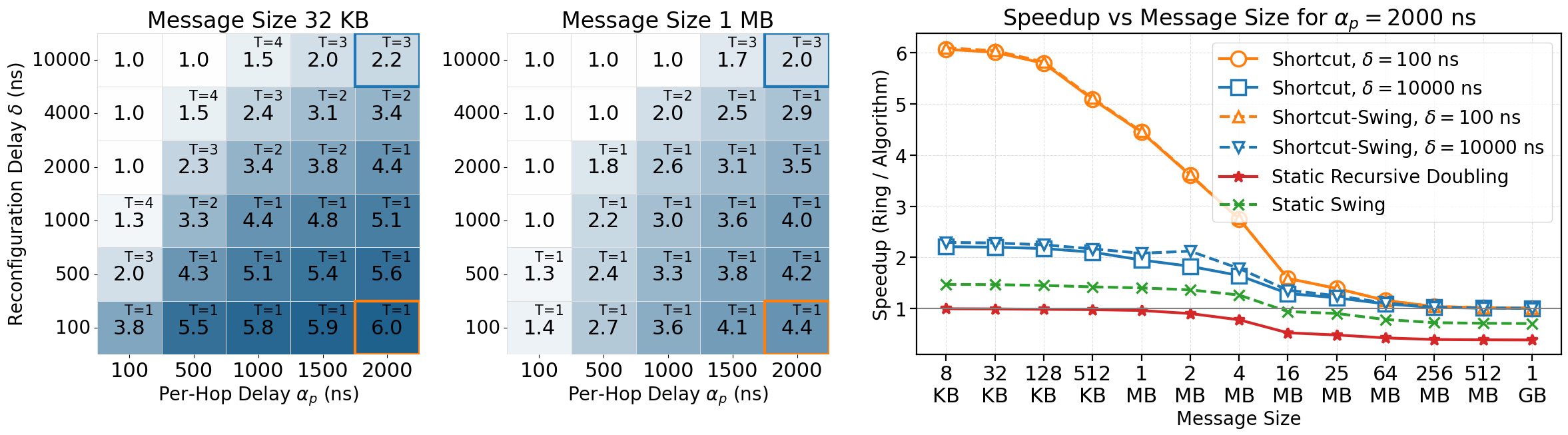}
\caption{Speedup and reconfiguration threshold $T$ for 32 nodes across message sizes. \emph{Left:} 32~KB. \emph{Middle:} 1~MB: the threshold shifts to $T=1$ as bandwidth cost grows. \emph{Right:} Speedup vs. message size at 2000~ns per-hop delay; hollow markers for \Shortcut with reconfiguration, filled markers for algorithms on a static ring topology.}
\label{fig:heatmaps}
\end{figure*}

\subsection{Methodology}\label{sec:methodology}

\sloppy
Our evaluations use Astra-Sim~\cite{rashidi2020astrasim} with ns-3 for packet-level simulation, which we extend to simulate reconfigurable circuit switches and direct-connect topologies. We sweep congestion control algorithms (DCQCN, HPCC, TIMELY), link bandwidths (400 and 800~Gbps), and MTU sizes (1500--9000~B), and report results for the configuration that appears least favorable to our approach: 800~Gbps links, 9000~B MTU, DCQCN. Evaluated reconfiguration delays span 100~ns end-to-end delay~\cite{ballani_Sirius_2020} to 10~$\mu$s~\cite{Multiport_farrington_2013}; per-hop delays span 100~ns, required for FEC in optical transceivers, to 2~$\mu$s~\cite{hu_demystifying_2025}.

We primarily evaluate ReduceScatter; AllGather is symmetric in terms of communication steps and AllReduce is their concatenation. For $n$ nodes, RD ReduceScatter takes $\log_2 n$ steps, resulting in $T$ values from $0$ (always reconfigure) to $\log_2 n$ (always static). Note that $T=0$ involves an unnecessary initial reconfiguration before the first step of ReduceScatter covered by the fixed matching, so $T=1$ is the first useful \enquote{always reconfigure} threshold, where topology is adjusted to perfectly match the communication pattern of each step. We report the reconfiguration threshold $T$ determined by Eq.~\ref{eq:reconfiguration_vs_ringreducescatter} and the speedup of RD using this threshold compared to the Ring algorithm on a fixed ring topology as $\frac{CCT_{\text{Ring}}}{CCT_{\text{\Shortcut}}}$, where CCT is collective completion time.

\subsection{Best Threshold for Recursive Doubling}
\label{sec:eval_best_strategy}

We first isolate the effect of reconfiguration on Recursive Doubling before comparing \Shortcut with other collective algorithms. This establishes when reconfiguration is useful at all and explains the schedules selected by \Shortcut in the subsequent comparisons.

We consider a topology of $n=32$ nodes and initially isolate network topology effects by setting $\alpha_s = 0$. Figure~\ref{fig:heatmap_best_T} shows the selected threshold for 32~KB ReduceScatter. A lower threshold means that \Shortcut begins reconfiguring earlier and therefore reconfigures more RD steps. As expected, \Shortcut selects lower thresholds when reconfiguration is faster, because the cost paid per reconfigured step decreases. It also selects lower thresholds as per-hop delay increases~--~this is where the gain of reconfiguration, to avoid cumulative multi-hop delays, is strongest. It selects lower thresholds because the communication distance of static RD grows exponentially with the step index. Consequently, the later RD steps are the first to become worth shortcutting; as reconfiguration becomes cheaper or multi-hop communication more costly, this profitable region expands toward earlier steps.

Message size affects the threshold through a different mechanism. Larger messages increase the cost of congestion created by overlapping multi-hop paths in static RD. As a result, avoiding congestion can justify reconfiguration even for earlier steps, shifting the selected threshold toward $T=1$, as shown by the annotations in Fig.~\ref{fig:heatmaps}~\emph{(left)} and \emph{(middle)}. For message sizes $\geq 4$MB, the threshold $T=1$~--~reconfiguring between every step~--~is selected all evaluated delays (not shown).

\subsection{\Shortcut vs. Static Algorithms}
\label{sec:eval_ReconfRD_vs_Ring}
We now compare the resulting \Shortcut schedule against two static-topology alternatives. Ring represents a topology-matched algorithm who naturally only communicates between neighbour nodes. The Swing algorithm~\cite{sensi_swing_2024} is a low-latency algorithm that adapts its communication pattern to reduce congestion on ring-based topologies, contrary to our approach, which instead adapts the topology to the algorithm. To ensure a fair comparison, nodes can send and receive one flow per step in all evaluated algorithms and baselines, as in~\cite{thakurOptimizationCollectiveCommunication2005}. We also apply the same threshold-based reconfiguration strategy of \Shortcut to Swing, using its step-specific path-length and congestion factors. Like RD, each Swing step induces a perfect matching and can therefore be realized directly by the reconfigurable fabric.

Interestingly, although the best reconfiguration threshold shifts earlier with increasing message size, this does not automatically translate to better performance relative to other algorithms on the static ring topology. Fig.~\ref{fig:heatmaps} illustrates these contrasting trends: as message sizes grow, the best-performing thresholds shift to earlier $T$ values (annotated), but the resulting speedup against the Ring algorithm decreases. At a small message size ($32$KB, Fig.~\ref{fig:heatmaps}~\emph{(left)}) our strategy is up to $6\times$ faster than the Ring algorithm but the same delay values result in a diminished speedup of $4.4\times$ when message sizes increase to $1MB$ in  Fig.~\ref{fig:heatmaps}~\emph{(middle)}.

Figure~\ref{fig:heatmaps}~(right) compares both reconfigured variants with their static counterparts. Static Swing improves over static RD by reducing topology-induced path penalties, but both remain slower than the \Shortcut variants for small and medium messages. \Shortcut-RD and \Shortcut-Swing achieve nearly identical performance because reconfiguration removes much of the costly path-length and congestion differences that distinguish the algorithms on a static topology. Their remaining costs are largely determined by their logarithmic step count and congestion-free transmission over direct links. This also shows that \Shortcut applies to other low-latency algorithms whose steps form perfect matchings. Both \Shortcut variants converge to the Ring algorithm baseline at large message sizes, and \Shortcut falls back to Ring for messages $\geq 256$~MB. This shows that the analytical formulation correctly captures the point at which the Ring algorithm is superior.

Topology reconfiguration delivers larger improvements than the static alternatives~--~especially with shorter reconfiguration $\delta$. The main advantage of reconfiguration over Ring is not congestion avoidance, since Ring also uses non-overlapping direct paths, but reduced cumulative delay. Reconfiguration allows low-step-count algorithms to retain their logarithmic latency by using direct links in the reconfigured steps.  At larger messages, bandwidth cost dominates total completion time, diminishing the relative impact of \Shortcut's latency savings. \Shortcut consequently provides its largest gains for small, latency-bound messages. 

\subsection{Step Count and Startup Latency}
\label{sec:eval_startup_delay}

\begin{figure*}[th]
\centering
\setlength{\tabcolsep}{6pt}
\renewcommand{\arraystretch}{1.0}
\begin{tabular}{ccc}
\includegraphics[width=0.29\textwidth]{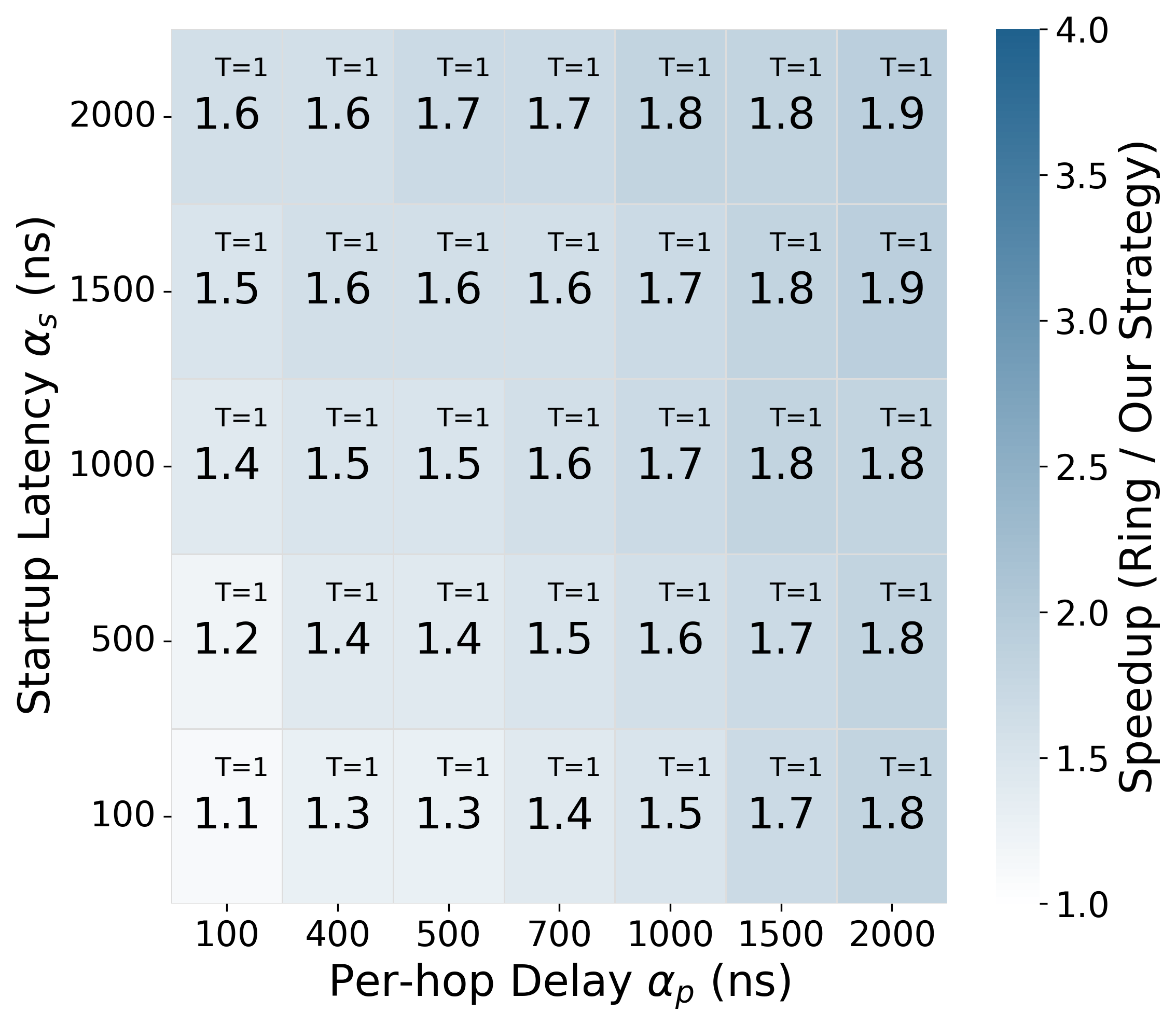} &
\includegraphics[width=0.29\textwidth]{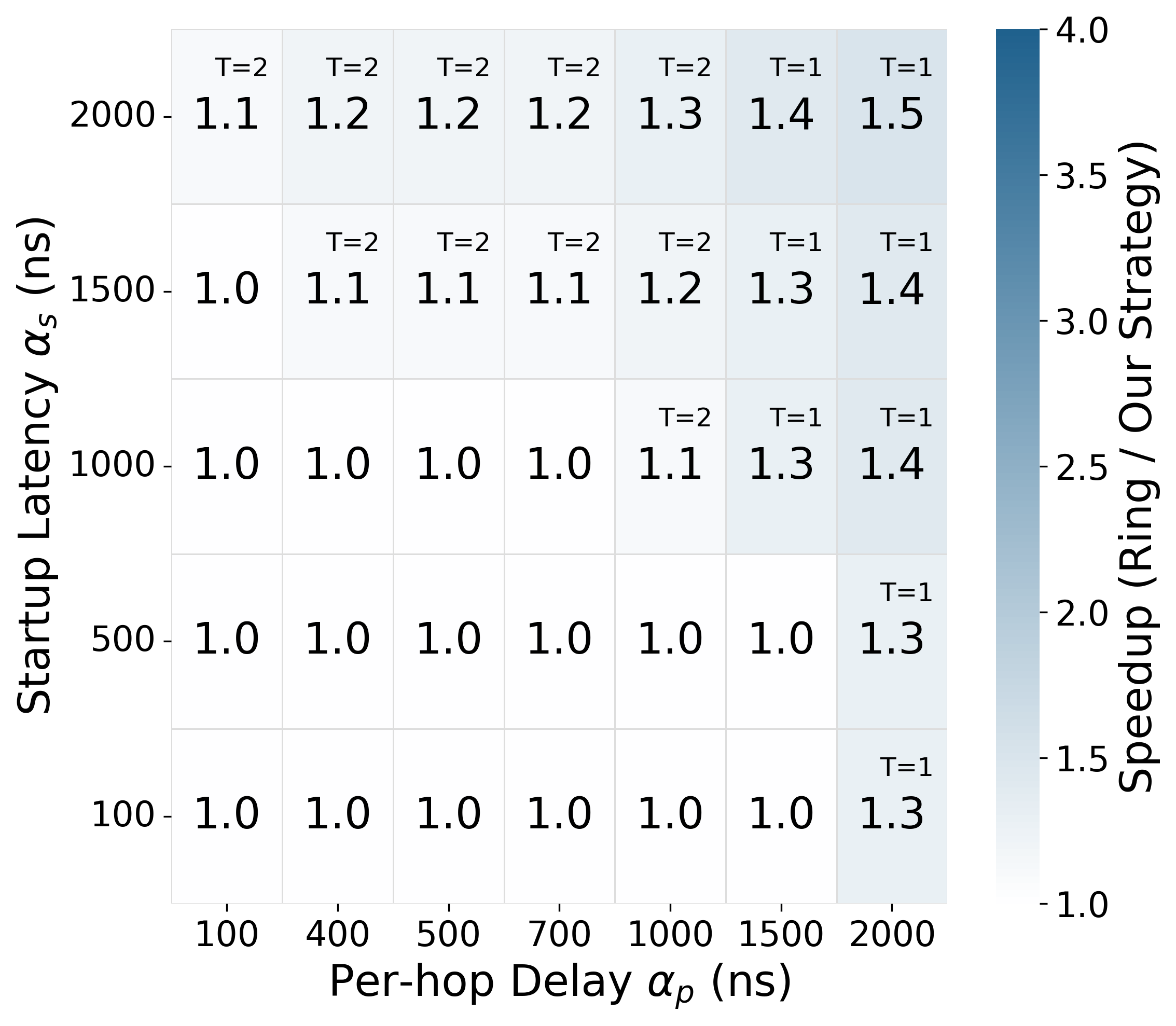} &
\includegraphics[width=0.29\textwidth]{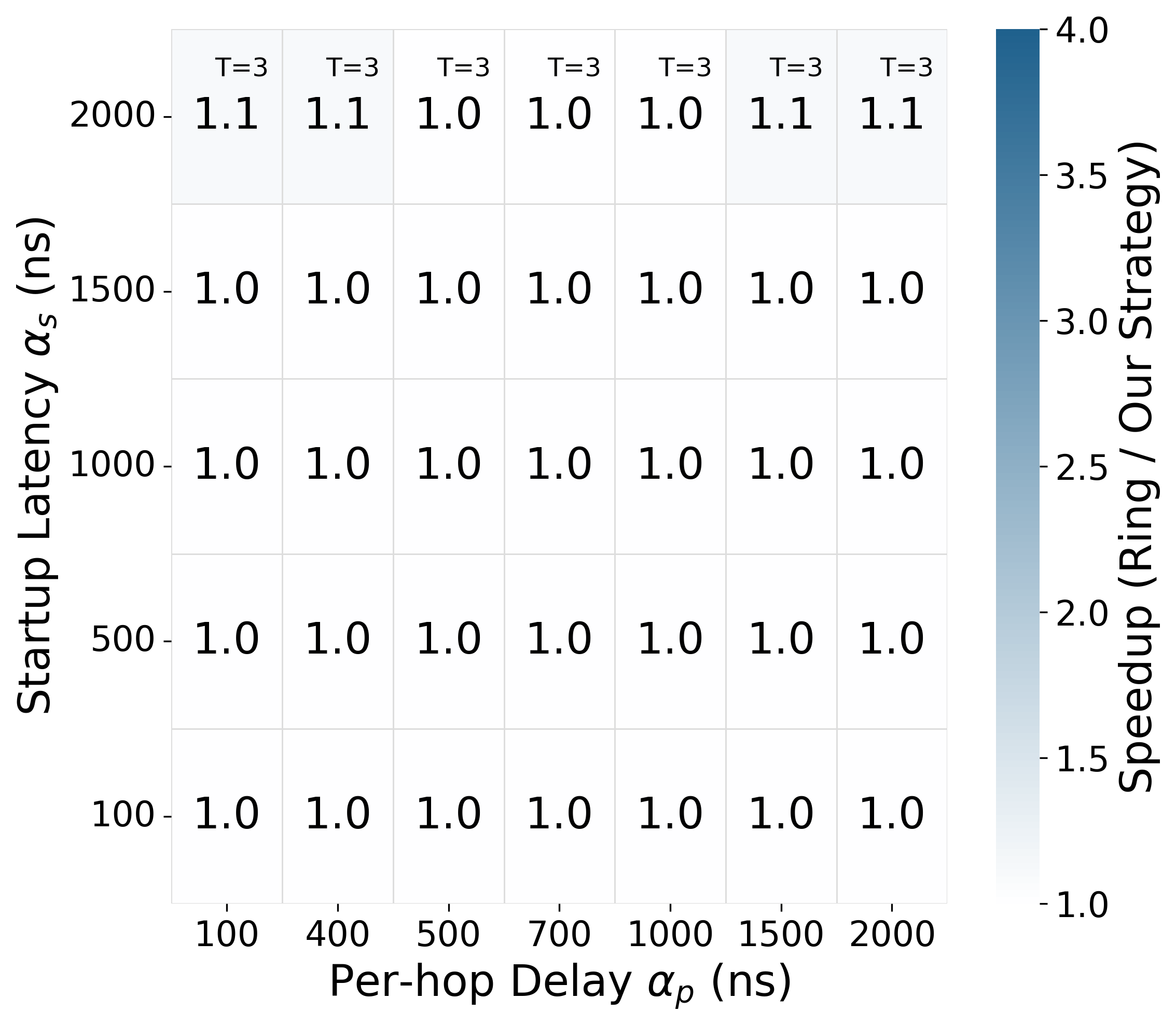} \\
\scriptsize (a) 8 nodes, $100$ns reconfig. &
\scriptsize (b) 8 nodes, $4\mu$s reconfig. &
\scriptsize (c) 8 nodes, $10\mu$s reconfig.\\
\includegraphics[width=0.29\textwidth]{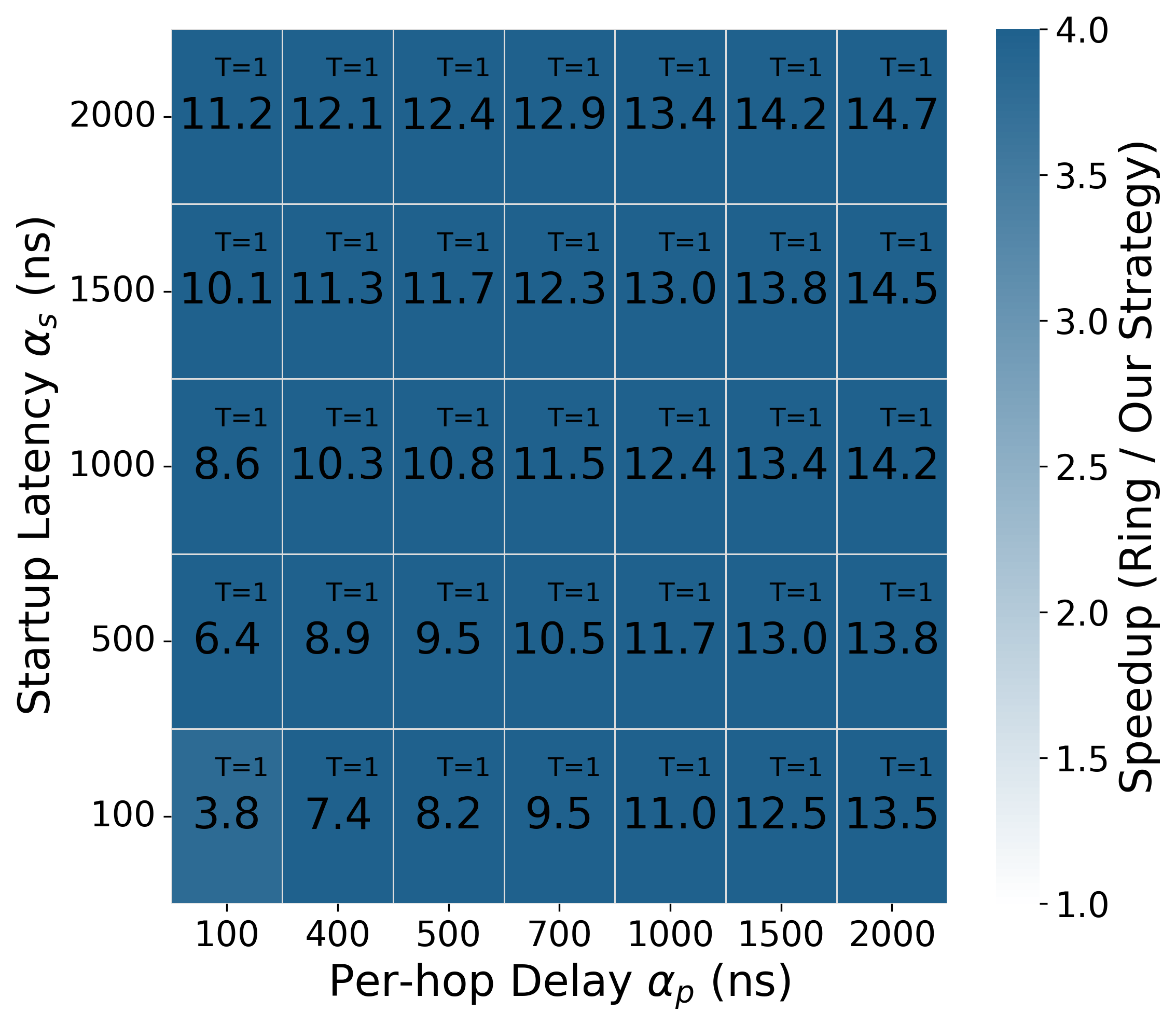} &
\includegraphics[width=0.29\textwidth]{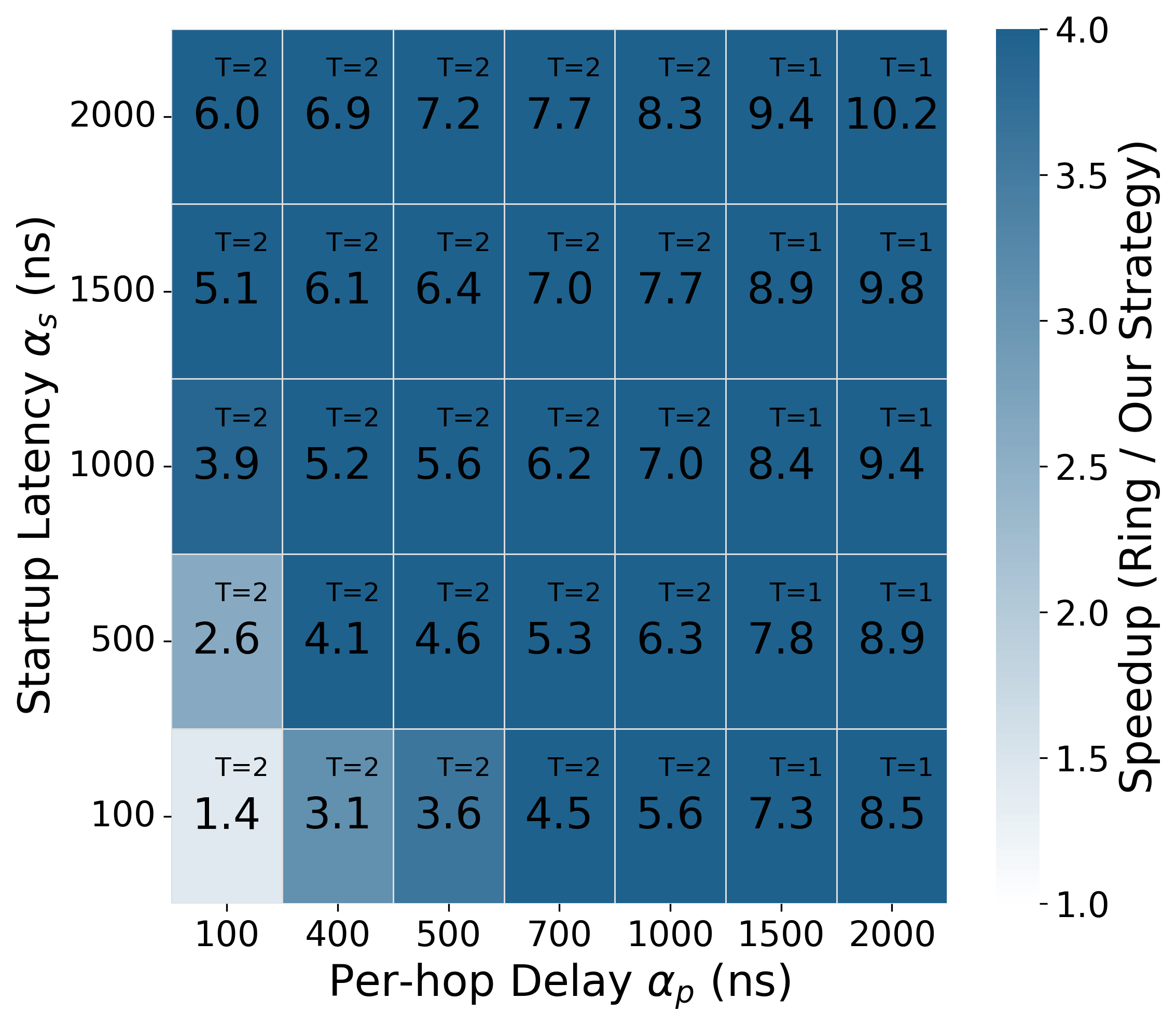} &
\includegraphics[width=0.29\textwidth]{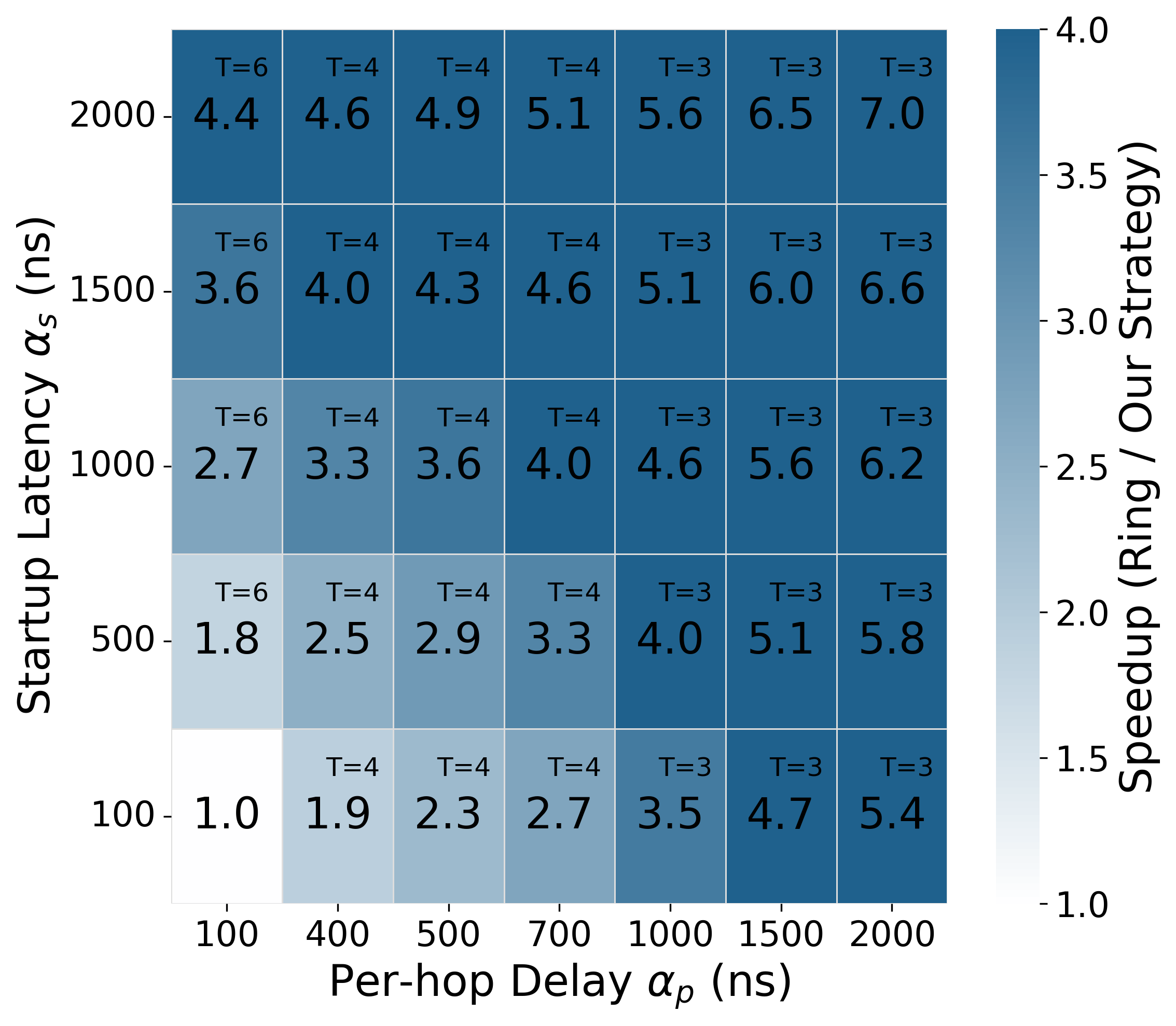} \\
\scriptsize (d) 128 nodes, $100$ns reconfig. &
\scriptsize (e) 128 nodes, $4\mu$s reconfig. &
\scriptsize (f) 128 nodes, $10\mu$s reconfig. 
\end{tabular}
\caption{Speedup of \Shortcut for 1~MB ReduceScatter relative to the static Ring baseline, varying startup latency $\alpha_s$ (y-axis) and per-hop delay $\alpha_p$ (x-axis), across reconfiguration delays (columns) and node counts (rows).}
\label{fig:startup_vs_prop_delay}
\vspace{-0.8em}
\end{figure*}

Startup latency is incurred once per step, so algorithm step count directly affects cumulative \emph{startup} latency.  We examine its interaction with propagation and reconfiguration delays at 1~MB, observed in large-scale deployments~\cite{RDMA_gangidi_2024} and latency-sensitive. We evaluate $\alpha_s$ up to 2~$\mu$s, covering reported values of 0.7~$\mu$s for NVLink and 1.7~$\mu$s for InfiniBand~\cite{285084}.

Panel~(c) of Fig.~\ref{fig:startup_vs_prop_delay} shows 8 nodes with $\delta{=}10\,\mu$s, where reconfiguration is too costly; \Shortcut selects $T{=}\log_2 n$, effectively comparing static RD and Ring. High startup latency slightly favors RD, but this is overshadowed by congestion, so \Shortcut falls back to Ring ($1\times$ speedup). When reconfiguration is feasible (panels a--b), substantially larger speedups emerge.  Reconfiguration extends RD's logarithmic advantage beyond startup latency to also cover per-hop delay: similar speedups appear at high $\alpha_s$ (top) and high $\alpha_p$ (right), since reconfiguration converts both into $O(\log n)$ costs.

\subsection{Impact of the Number of Nodes}
\label{sec:eval_node_number}
Data Parallelism, which uses on AllReduce to synchronize gradients in ML trainin, commonly spans 8--128 accelerators~\cite{RDMA_gangidi_2024,Llama_grattafiori_2024,Collective_si_2025}; we therefore evaluate both endpoints of this range. These ring sizes are representative on torus-based interconnects specialized for distributed ML training, such as TPUv4~\cite{jouppiTPUV4Optically2023}, where rings form the building blocks of higher-dimensional tori. Additionally, different parallelism dimensions are often mapped onto separate torus dimensions~\cite{jouppiTPUV4Optically2023}.

Our analysis suggests that larger node counts amplify \Shortcut's benefits. Ring's startup and propagation latencies both grow linearly with $n$, while its bandwidth term $\frac{n-1}{n}m\beta$ barely changes. This makes latency increasingly dominant in the Ring algorithm as $n$ grows. \Shortcut targets precisely this latency term: reconfiguration enables RD to maintain logarithmic cumulative delay even as $n$ increases, while RD on a static topology suffers linear multi-hop penalties that worsen with $n$. Fig.~\ref{fig:startup_vs_prop_delay} reflects this: the 128-node collective (bottom row) achieves speedups up to $\sim 14.7\times$ and significantly benefits from reconfiguration even at $\delta{=}10\,\mu$s (panel~f), compared to modest gains at 8 nodes in the row above.

The 100~ns lower bound evaluated here is among the fastest demonstrated OCS technologies~\cite{ballani_Sirius_2020}, but achieved only in small-scale prototypes. Photonic MEMS switches have been demonstrated with 240 ports and 400~ns reconfiguration delay~\cite{Waferscale_seok_2019}, placing it between panels~(d) and~(e) and suggesting an order-of-magnitude improvement remains within reach.
 
\section{Discussion}\label{sec:discussion}

\textbf{Collective communication algorithm synthesis}, such as TACCL~\cite{285084} and TE-CCL~\cite{10.1145/3651890.3672249}, derives optimized schedules by exploiting structural diversity in the network topology. 
The uniform, low-degree ring topology studied here offers limited such diversity, hence few opportunities for pure algorithm synthesis. Optical circuit switching introduces new connectivity options; co-optimizing topology and communication scheduling during collectives is an exciting future direction.

\textbf{Optical Interconnect.} The schedule calculation is simple and local by design: all inputs are either static (e.g., link bandwidth $\beta$) or locally derivable from the workload (e.g., message size $m$), the $O(\log n)$ candidate thresholds can be evaluated without coordination, and the complete reconfiguration sequence is determined before the collective begins. This maps naturally onto OCS designs with passive cores, where nodes can independently calculate $T$ and steer light via laser wavelength tuning~\cite{ballani_Sirius_2020}. Since the per-step connectivity is fully known in advance and the same collective repeats across ML training iterations, the receiver can apply phase caching~\cite{ballani_Sirius_2020} to achieve sub-nanosecond lock-in at every reconfiguration.

\textbf{Other Collective Algorithms.} \Shortcut targets pairwise collective algorithms in which each step induces a symmetric matching. This allows the communication pattern to be realized with a single dynamic OCS port per node. The same principle can be extended to collective algorithms with antisymmetric matchings such as Bruck, which however require additional optical ports per node. These architectural requirements may increase cost and reconfiguration delay. We therefore focus on the minimal configuration with one dynamic optical port per node.

\textbf{Synchronization Between Reconfigurations.} An OCS can be reconfigured only after all transmissions using the current topology have completed; otherwise, active traffic may be disrupted and data lost. Reconfiguring before step $k$ therefore requires completion of all transmissions in step $k-1$, followed by synchronization among the participating nodes. Barrier overhead, stragglers, and control-plane latency may consequently reduce the benefit of frequent topology updates. We account for these effects by evaluating effective reconfiguration delays that exceed the switching times reported for selected OCS prototypes.

\section{Conclusion}
\label{sec:conclusion}

We believe \emph{in-collective} reconfiguration is an opportunity that should not be missed, especially as the reconfiguration delays of optical switching continue to decrease with technological advances. This paper offers an understanding of the underlying trade-offs between important low-latency and topology-matched AllReduce algorithms and a glimpse of the performance gains possible.

\section{Use of AI}
ChatGPT 5.6 was used to assist with editing and provide partial code to plot results. Ideas and content remain original.

\bibliographystyle{IEEEtran}
\bibliography{references}

\end{document}